# Thermal and chaotic distributions of plasma in laser driven Coulomb explosions of deuterium clusters

M. Barbarino[1], M. Warrens[1,2], A. Bonasera[1,3], D. Lattuada[1,3,4], W. Bang[5], H. J. Quevedo[6], F. Consoli[7], R. De Angelis[7], P. Andreoli[7], S. Kimura[8], G. Dyer[6], A. C. Bernstein[6], K. Hagel[1], M. Barbui[1], K. Schmidt[1], E. Gaul[6], M. E. Donovan[6], J. B. Natowitz[1], and T. Ditmire[6]

[1]*Cyclotron Institute, Texas A&M University, College Station, Texas 77843, USA*
[2]*University of Dallas, Irving, TX 75062*
[3]*LNS-INFN, via S. Sofia, 62, 95123 Catania, Italy*
[4]*Universita' degli studi di Enna "Kore", 94100 Enna, Italy.*
[5]*Los Alamos National Laboratory, Los Alamos, New Mexico 87545, USA*
[6]*Center for High Energy Density Science, C1510,
University of Texas at Austin, Austin, Texas 78712, USA*
[7]*Associazione Euratom-ENEA sulla Fusione, via Enrico Fermi 45, CP 65-00044 Frascati, Rome, Italy and*
[8]*Department of Physics, University of Milano, via Celoria 16, 20133 Milano, Italy*
(Dated:)

In this work we explore the possibility that the motion of the deuterium ions emitted from Coulomb cluster explosions is chaotic enough to resemble thermalization. We analyze the process of nuclear fusion reactions driven by laser-cluster interactions in experiments conducted at the Texas Petawatt laser facility using a mixture of $D_2$+$^3$He and $CD_4$+$^3$He cluster targets. When clusters explode by Coulomb repulsion, the emission of the energetic ions is "nearly" isotropic. In the framework of cluster Coulomb explosions, we analyze the energy distributions of the ions using a Maxwell-Boltzmann (MB) distribution, a shifted MB distribution (sMB) and the energy distribution derived from a log-normal (LN) size distribution of clusters. We show that the first two distributions reproduce well the experimentally measured ion energy distributions and the number of fusions from d-d and d-$^3$He reactions. The LN distribution is a good representation of the ion kinetic energy distribution well up to high momenta where the noise becomes dominant, but overestimates both the neutron and the proton yields. If the parameters of the LN distributions are chosen to reproduce the fusion yields correctly, the experimentally measured high energy ion spectrum is not well represented. We conclude that the ion kinetic energy distribution is highly chaotic and practically not distinguishable from a thermalized one.

## I. INTRODUCTION

Experiments using intense lasers irradiating clusters are interesting not only for practical application such as energy production and neutron sources but also for basic science such as the measurement of cross sections in hot plasma of interest in astrophysics. In such studies [1–23], the interaction between the laser pulse and the clusters causes the ionization and subsequent explosion of the clusters creating a hot plasma. Assuming that all the electrons are stripped by the laser light, the explosion and fusion of the cluster targets is well described by the Coulomb explosion model [1–4, 10, 13, 14, 16, 24–30] in which the energetic ions resulting from the laser-target interaction accelerate and fuse within the plasma. It is the collisions among ions of different clusters which gives rise to nuclear fusion. Since the fusion cross sections for d-d and d-$^3$He reactions are very sensitive to the high energy tails of the plasma ion distributions, it is very important to determine the true nature of the particle energy distribution functions and perform the data extrapolation precisely. During the experiments [3, 4, 30], an intense ultra-short laser pulse irradiates either $D_2$ or $CD_4$ clusters mixed with $^3$He gas, simultaneously producing three types of nuclear fusion reactions in the interaction volume: D(d, t)p, D(d, $^3$He)n and $^3$He(d, p)$^4$He. In this work, we focus on the study of the last two reactions, from which 2.45 MeV neutrons and 14.7 MeV protons are produced, respectively. According to the Coulomb explosion mechanism, the kinetic energies of the resulting deuterium ions reach several keV and d-d fusion reactions can occur when energetic deuterium ions collide with each other or with cold deuterium atoms in the background gas jet outside the focal spot, see also [31]. For d-$^3$He fusion reactions, the $^3$He ions are regarded as stationary since they remain cold after the intense laser pulse is gone (i.e., $^3$He atoms do not absorb the laser pulse energy efficiently because they do not form clusters at the nozzle temperature of 86 K [30, 32]). In this paper, we analyze both fusion yields of neutrons and protons produced in three different scenarios of ion energy distributions (i.e., Maxwell-Boltzmann (MB), shifted Maxwell-Boltzmann (sMB) and log-normal (LN) distributions). We show that the fusion yields and the plasma ion kinetic energy distributions are consistent with MB and sMB. The LN distribution does not offer a good representation of the measured ion signal when reproducing the measured fusion yields. Thus we conclude that, at least for the quantities considered here, the system is close to a thermal state distribution scenario. It is important to stress that equilibrium is probably not reached through ion-ion, ion-electron etc. collisions since the ion kinetic energies are of the order of keV or higher. But rather, the Coulomb explosion of different cluster sizes is so chaotic as to be practically not distinguishable from thermalization [33].

## II. EXPERIMENTAL SETUP

The Texas Petawatt laser (TPW) is a 190 J, 170 fs laser based on Optical Parametric Chirped Pulse Amplification followed by power amplification in two types of Neodymium-doped glass [34]. During the experiments, the TPW delivered 90-180 J per pulse with 150-270 fs duration onto a cryo-cooled gas mixture of $D_2+{}^3$He or $CD_4+{}^3$He released from the gas jet [32]. The intense laser beam that irradiates the clusters removes the electrons from the atoms and causes the clusters of deuterium ions to explode by Coulomb repulsion creating a hot plasma. Nuclear reaction occurring between the ions within a single cluster is negligible compared with reaction between ions belonging to different clusters [1, 2]. An f/40 focusing mirror (10 m focal length) created a large interaction volume in this experiment with laser intensities sufficiently high to drive laser-cluster fusion reactions. The radius, $r$, of the cylindrical fusion plasma was estimated from the beam profile measured at the equivalent image plane of the cluster target. Two cameras imaged the side and bottom of the plasma on each shot. The ratio of the atomic number densities of deuterium and ${}^3$He for each shot was calculated from a residual gas analyzer which measured the partial pressures of $D_2$, $CD_4$, and ${}^3$He in the mixture. The gas mixtures were introduced at a pressure of 52.5 bars into a conical supersonic nozzle with a throat diameter of 790 $\mu$m, an exit radius $R$ of 2.5 mm, and a half angle of 5° to generate large clusters (diameter >10 nm) necessary for energetic cluster explosions. $D_2+{}^3$He and $CD_4+{}^3$He mixtures were cooled to 86 K and 200-260 K, respectively, to maximize the production of large clusters. Five calibrated plastic scintillation detectors [35] measured the yield of 2.45 MeV neutrons generated from d-d fusion reactions. Three plastic scintillation detectors measured the yield of 14.7 MeV protons from the ${}^3$He(d,p)${}^4$He fusion reactions. These proton detectors were located in vacuum 1.06-1.20 m from the plasma at 45°, 90°, and 135°. Ion energy distribution signals were measured using a Faraday cup of 16 mm diameter, placed at 157.5° relative to the laser direction at a distance of about 1.07 m from the target, with a -400 V bias [3, 4, 30]. By measuring the Faraday cup signal ($\Delta V$), one can determine the number of ions $N$ hitting the cup in energy space $E$ as [11]

$$\frac{d^2N}{dEd\Omega} = \frac{s^3}{m_D v^3 \pi r_F^2} \frac{\Delta V}{qeR_\Omega}. \quad (1)$$

In the above equation, $s$=1.07 m is the distance of the detector from the target, $m_D$ and $v$ are the deuterium ion mass and speed, respectively, $q$ is the charge state (i.e., $q$=1 for deuterium), $e$ is the electron elementary charge (i.e., $e$=1.6×10${}^{-19}$ C), $R_\Omega$=50 $\Omega$ is the Faraday cup impedance, $d\Omega = \frac{\pi r_F}{s^2}$ is the solid angle and $r_F$=8 mm is the radius of the Faraday cup detector. We will assume that the ion angular distribution is flat [1]. In order to distinguish the detectable ion signal from the electromagnetic pulse (EMP) and X-ray and highlight these different structures, the system can be studied by multiplying the energy distribution by $E^n$ (i.e., $n$=0,1,2) [11]. In Fig. 1, we show some typical kinetic energy distributions of deuterium ions obtained using a mixture of $D_2$ and ${}^3$He gases.

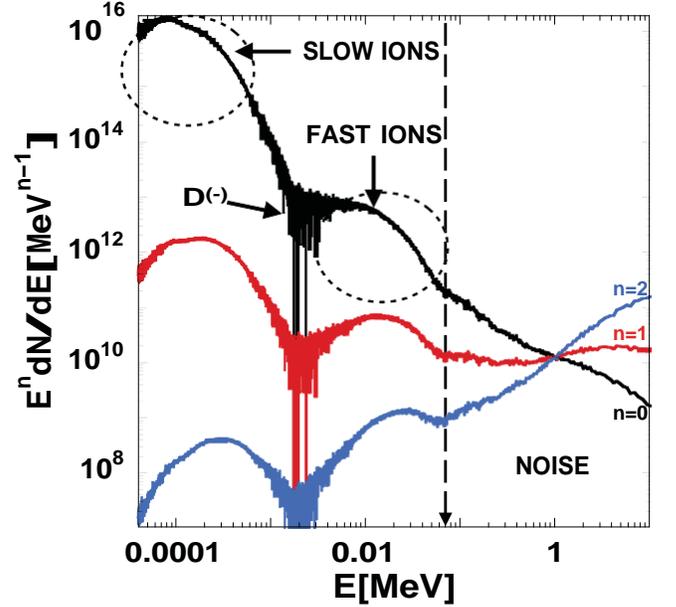

FIG. 1. Ion kinetic energy moments distribution $n$=0 (black line), $n$=1 (red line) and $n$=2 (blue line) of the signals for a shot performed using a mixture of $D_2+{}^3$He. Moments analysis is a powerful tool to separate the electromagnetic noise from detectable signal.

Three regions were identified. In the low energy region below 1 keV, the ion signal belongs to the blast wave of the energetic plasma ions [1]. In the intermediate region below roughly 70 keV, the ion signal represents the kinetic energy spectrum of deuterium ions coming from the Coulomb explosion of large clusters, these are the ions responsible for fusion reactions, especially d-d. d-${}^3$He fusion reactions are mostly generated from ions very close to where the noise becomes dominant. In the high energy region starting from roughly 70 keV, the ion signal overlaps with the EMP and the X-ray "noise". From the kinetic energy moments distribution analysis it is relatively easy to separate these three regions and distinguish the electromagnetic noise from detectable signal. Notice the large oscillations in the distributions around 1 keV. Those oscillations are not due to the initial EMP nor to the X-ray, since they are detected after a relatively long time (>2 $\mu$s). We could assume that the energetic ions coming from the laser-cluster interaction region quickly expand into the cold region of the plasma (i.e., the part that was not irradiated by the plasma). In their path, they capture some electrons which are still in the surrounding medium and especially the lower energy ions might become negatively charged. Thus, it is the net sum of positive and negative charges which produces the oscillations. Negative ions of kinetic energies lower than 400 eV are rejected by the repulsive grid in the Faraday cup which terminates the oscillations. However, since such an energy region is irrelevant for the fusion reactions to occur, this feature is not discussed further in this work. Note that the ion signal roughly above > 70 keV, where the fusion cross-sections are large, is obscured by the noise. It is important to stress that, if relevant information is to be determined from these experiments, such as fusion cross sections [4, 11, 23, 36], it is crucial to have a clean signal also above these energies interval. It is the same energy region that, as we will show in this paper, is absolutely needed to be able to distinguish the MB distributions from the LN. Because of the noise shown in Fig. 1, we analyzed the number of fusions to pin down the most suitable distribution. We calculated the total number and energy spectrum of deuterium ions in the plasma, with the assumption of nearly isotropic emission [37, 38]. This assumption is reasonable because the clusters undergo Coulomb explosion [26]. In future experiments, further improvements must be made in the detection of the plasma ion signal and the precision of the fusion particle yields.

## III. MAXWELL BOLTZMANN DISTRIBUTION: THERMAL STATE

Assuming that all the electrons are stripped by the laser light, the 'naked' clusters will quickly explode because of the Coulomb repulsion among the ions and the kinetic energy will result in the measured temperature $T$. We assume that the ion kinetic energy distributions can be described by either Maxwell-Boltzmann [3, 4, 25] or shifted Maxwell-Boltzmann distributions (i.e., $E \rightarrow E - E_C$) [39]. In the second scenario, we find that the ions go through a cloud of electrons and experience a positive attraction or energy loss causing a shift in the energy distribution [11]. The kinetic energy moments distribution of the deuterium ions for charge state $q=1$ can be written as [11]

$$E^n \frac{dN}{dE} = \frac{N_0}{\sqrt{\pi T^3}} E^n (E - E_C)^{\frac{1}{2}} \exp\left(-\frac{(E - E_C)}{T}\right), \quad (2)$$

where $n=0,1,2$ and $N_0$ is a normalization constant. The parameters in Eq. (2) are fitted to the experimentally measured ion energy distributions, (see Fig. 2, $n=0$). Two distributions are used for each energy region, the low and the intermediate region respectively. Note that on the shots examined, the shifted MB distribution for the intermediate energy region gives an average negative $E_C \sim -11$ keV. This could be consistent with a deceleration due to a cloud of electrons still surrounding the exploding clusters causing a total positive shift in the energy because of the attractive collective energy (i.e., negative potential) experienced by the ions.

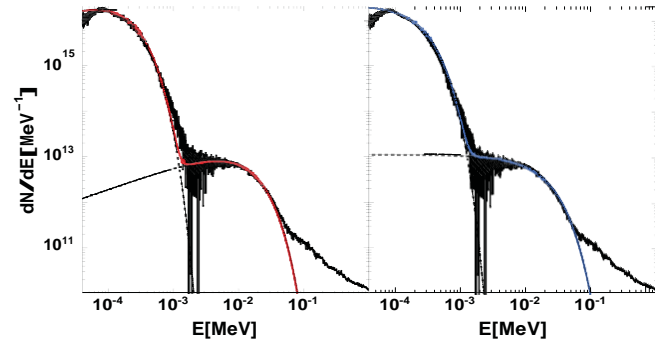

FIG. 2. Ion kinetic energy distributions obtained using a Maxwell-Boltzmann distribution $E_C=0$ (left panel) or a shifted Maxwell-Boltzmann distribution $E_C \neq 0$ (right panel) for a shot performed using a mixture of $D_2+^3He$. The dashed lines correspond to a fit of each ion region (i.e., fast ions and slow ions) whose sum gives the total contribution (solid line).

It is important to note that the fusion plasmas might not be in thermal equilibrium even though the Maxwell-Boltzmann distribution reproduces the experimentally measured ion energy spectrum rather well. The consistency of this distribution scenario also remains when introducing a collective energy. However, the energetic deuterium ions in the plasma have such high kinetic energies that their mean free paths (>10 mm) are longer than the size of the gas jet, and thermalization from ion-ion collisions or ion-electrons is not expected under our experimental conditions. Previous studies have in fact shown that near-Maxwellian ion energy distribution is observed not because of thermalization of ions, but because of cluster size distribution [29, 37]. In the following section, we consider this possibility that a near-Maxwellian energy distribution results from the log-normal size distribution of clusters in our gas jet.

## IV. LOG-NORMAL DISTRIBUTION: CHAOTIC STATE

In this section we will explore the possibility that the motion of the ions within a cluster is chaotic enough to resemble thermalization. This is typical of many bodies interacting through long range forces which give rise to chaos already for a 3 particle system [33]. Microscopic simulations of such processes show that fusion might indeed occur even though a collective motion is initially imposed to the system [40]. Here we do not suggest that it is either one mechanism or the other, but simply we want to explore a different scenario and maybe suggest how to assess their relative contributions exper-imentally. Previous studies of cluster production in gas jets have found that the experimentally measured size distribution is best described by a log-normal distribution [3, 19, 20, 25, 37, 38, 41, 42]. In that case the clusters of size $M$ density moments can be written as [37]

$$M^n \frac{dN}{dM} = \frac{M_0 M^{n-1}}{\sqrt{2\pi\sigma^2}} \exp\left(-\frac{(\ln M - \mu)^2}{2\sigma^2}\right). \quad (3)$$

Where $n=0,1,2$. In the above equation, $M_0$ is a normalization constant, $\mu$ and $\sigma$ are the mean and the standard deviation of the distribution of the natural logarithm of the size. Assuming the relation between the radius and the number of ions $M$ in the cluster is

$$R_{cl} = r_s M^{\frac{1}{3}}, \quad (4)$$

where $r_s = (\frac{3}{4\pi\rho_{cl}})^{\frac{1}{3}} = 1.7 \text{Å}$ [28], then the Coulomb energy $V_C$ per particle for a uniformly charged sphere of radius $R_{cl}$ can be written as [2]

$$\frac{V_C}{M} = 5.1 M^{\frac{2}{3}} (eV) \equiv E_d. \quad (5)$$

Thus we are assuming that the kinetic energy of a deuterium ion is due to the Coulomb explosion of a cluster of size M. Different cluster sizes result in different ion energies. The relation between the quantities $E_d$ and $M$ can therefore be estimated from the equation above (i.e., $\frac{dE}{dM} = 3.4 M^{-\frac{1}{3}}$) and the Maxwell-Boltzmann function (i.e., $n=0$) can be rewritten in terms of $M$ as ($\frac{dN}{dM} = \frac{dN}{dE}\frac{dE}{dM}$) [11]

$$\frac{dN}{dM} = \frac{3.4 N_0}{M^{\frac{1}{3}} \sqrt{\pi T^3}} ((5.1 M^{\frac{2}{3}}) - E_C)^{\frac{1}{2}}$$
$$\exp\left(-\frac{((5.1 M^{\frac{2}{3}}) - E_C)}{T}\right). \quad (6)$$

This allows us to analyze the cluster size distribution in terms of a "pseudo" Maxwell distribution. Alternatively, from the log-normal distribution function, we obtain the ion kinetic energy distribution function (i.e., $n=0$) using the same relationships as [11]

$$\frac{dN}{dE} = \frac{1.4 M_0}{E\sqrt{2\pi\sigma^2}} \exp\left(-\frac{(\ln(0.09 E^{\frac{3}{2}}) - \mu)^2}{2\sigma^2}\right). \quad (7)$$

Notice that an alternative derivation to the above result is discussed in [2]. Even though the two approaches seem to present different functional terms, the logarithm dependence of the energy produce similar results at high energies and therefore we will not discuss further the latter in this paper. These sets of equations can be used to study the plasma ion kinetic energy moments distribution via Eqs. (2) and (7) or the cluster size moments distribution produced using Eqs. (3) and (6), see Fig. 3.

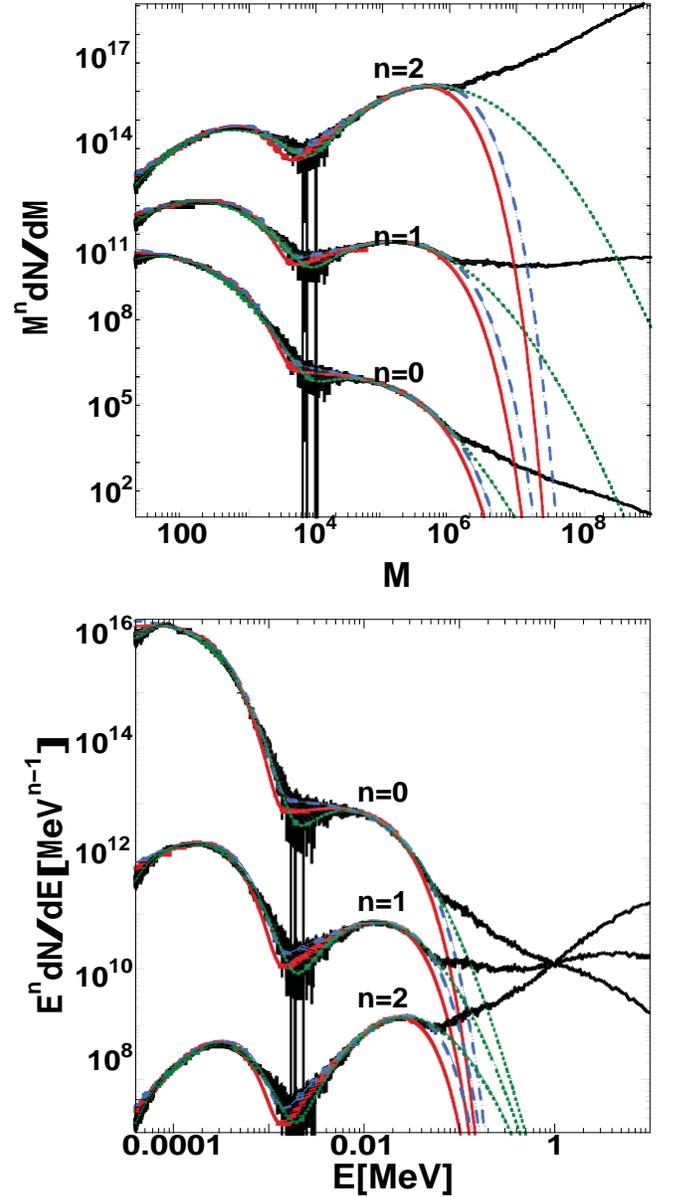

FIG. 3. Moments analysis ($n=0,1,2$) of the signal both in clusters space (top panel) via Eqs. (3) and (7) and energy space (bottom panel) via Eqs. (2) and (7). Maxwell-Boltzmann distribution (solid red), shifted Maxwell-Boltzmann distribution (dashed blue) and log-normal distribution (dotted green) are plotted. Note especially the differences for large cluster sizes. Thus a precise measurement of the high energy ions is crucial to distinguish the different distribution. Alternatively a calculation of the number of fusions can exclude some of the hypothesized distributions (or all). For higher moments ($n=1,2$), the noise is clearly visible and it occurs exactly where the distributions differ most.

In Eqs. (3) and (7), the presence of the natural logarithm makes the functions decrease more slowly at higher energy and larger cluster size (i.e., log-normal distribution goes to zero slower for energy or cluster size approaching infinity). This is shown in Fig. 3 where discrepancies among the different distributions are es-pecially visible for high energies (> 70 keV) and large cluster sizes. For this reason, the fact that despite all distributions are able to describe the measured distribution quite well, the substantial difference these show in the high momentum tail of the signal will be dominant when calculating the fusion yield. This analysis might give us a hint as to which one is the process that governs the energy distribution of the plasma we measure, and perhaps show us the true nature of the signals recorded. From Fig. 3, we could conclude that all distributions reproduce the data relatively well up to the region where the noise becomes dominant. Thus we need further information to distinguish them. This might be accomplished by comparing the number of fusions obtainable from the fitted distributions with the measured particle yield. A closer look at the energy distribution of the plasma in energy space is given in Fig. 4.

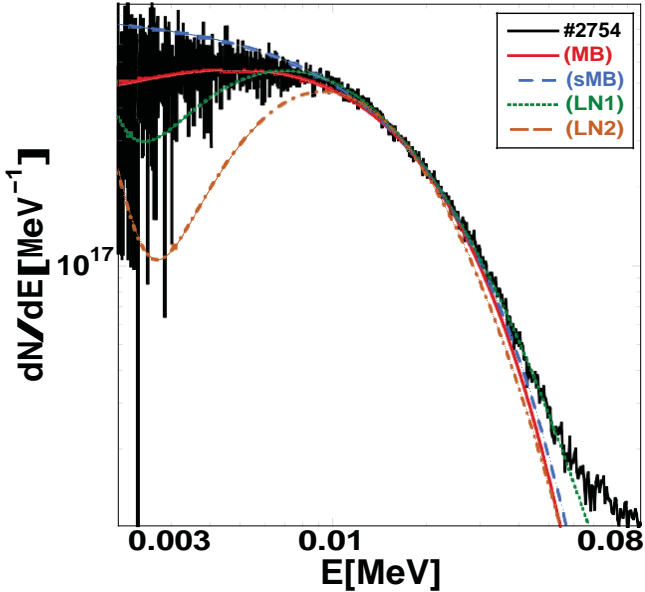

FIG. 4. Comparison between different plasma ion kinetic energy distributions (n=0) in the intermediate energy region. MB distribution (solid red line), shifted MB distribution (dashed blue line) and log-normal distributions LN1 (dotted green) and LN2 (dash-dot orange line) are plotted. The plasma ion kinetic energy distributions and the fusion yields are consistent with MB and sMB. The LN distribution does not reproduce both the measured ion signal and the fusion particle yields with the same set of parameters. In particular, the dotted green line (LN1) reproduces the measured distribution quite well but not the fusion yield, and vice versa the dash-dot orange line (LN2). Only the ions relevant to fusion reactions are plotted in the figure.

## V. FUSION REACTIONS

As a matter of fact, each one of the particle distributions described will determine a different fusion cross section and to test our distributions we estimate the number of fusion events under each different scenario. In general, the total fusion yield produced via laser-plasma reactions can be estimated as [1, 3, 4, 11]

$$Y = \frac{\rho_1 N_2 \langle \sigma v \tau \rangle}{1 + \delta_{12}}. \quad (8)$$

Where $\rho_1$ is the target density of species 1, $N_2$ is the total number of laser accelerated ions of species 2, $\langle \sigma v \rangle$ is the average reactivity, $\tau$ is the plasma disassembly time. The value of $\delta_{12}$ in the denominator is "0" for the case where the two species of particles are different and "1" for the case where they are the same. Similarly to [3, 4, 30], we focus on the study of the reactions from which 2.45 MeV neutron ($Y_n$) and 14.7 MeV proton ($Y_p$) yields are produced, respectively. As previously discussed, the kinetic energies of the deuterium ions resulting from the Coulomb explosion mechanism reach several keV so that d-d fusion reactions can be generated when energetic deuterium ions collide with each other called beam-beam fusion ($Y_{n(BB)}$) or with cold deuterium atoms in the background gas jet outside the focal spot called beam-target fusion ($Y_{n(BT)}$). Similarly to [3, 4, 30, 43], we will estimate the probability of d-d fusion in the cluster plume in the limit where the plasma disassembly time for collisions involving hot deuterium ions only (BB) can be approximated as [11]

$$\tau_{BB} = \frac{l}{v}. \quad (9)$$

In the above equation, $l$ is the radius of a sphere with volume equal to a cylindrical plasma of radius, $r$, and height, $R$, and $v$ is the speed of the hot deuterium ions. On the other hand, for collisions between hot deuterium ions with cold deuterium atoms (BT), the plasma disassembly time can be estimated as [11]

$$\tau_{BT} = \frac{R-l}{v}. \quad (10)$$

Thus, we consider only the region outside the fusion plasma, over a distance $(R-l)$. Whereas the probability of d-$^3$He fusions in the cluster plume is estimated in the limit where the plasma disassembly time is determined as [11]

$$\tau_{d^3He} = \frac{R}{v}. \quad (11)$$

Therefore, the 2.45 MeV neutron yield $Y_n$ is calculated as [3, 4, 11, 30]

$$Y_n = Y_{n(BB)} + Y_{n(BT)} = \frac{\rho_D N \langle \sigma \rangle_{dd_{(BB)}} l}{2} + \rho_D N \langle \sigma \rangle_{dd_{(BT)}} (R-l). \quad (12)$$

where $N$ is the total number of energetic deuterium ions in the plasma, $\rho_D$ is the average atomic number density of deuterium cluster plume, $\langle \sigma \rangle_{dd_{(BB)}}$ is the average fusion cross section between hot deuterium ions, $\langle \sigma \rangle_{dd_{(BT)}}$ is the average fusion cross section between hot deuterium ions and cold deuterium atoms at $\frac{T}{2}$ since one of the ions is cold (i.e., $E^{dd_{(BT)}}_{c.m.} = \frac{1}{2} E^{dd_{(BB)}}_{c.m.}$) [3, 4, 11, 30].

On the other hand, the 14.7 MeV proton yield is calculated as [3, 4, 11, 30]

$$Y_p = \rho_{^3He} N <\sigma>_{d^3He} R, \quad (13)$$

where $\rho_{^3He}$ is the average atomic number density of $^3$He and $<\sigma>_{d^3He}$ is the average fusion cross section between hot deuterium ions and cold $^3$He ions at $\frac{3}{5}T$ since $^3$He is at rest (i.e., $E^{d^3He}_{c.m.} = \frac{3}{5} E^{dd_{(BB)}}_{c.m.}$) [3, 4, 11, 30]. In our calculations, we make use of the deuterium cluster density and $^3$He concentration measured during each shot. Then, to determine which ion kinetic energy distribution best reproduces the experimental yields, the average fusion cross section of each reaction is estimated numerically or analytically from each energy distribution scenario as [11]

$$N<\sigma> = \int_0^\infty \sigma(E) \frac{dN}{dE} dE. \quad (14)$$

In the above equation, $\sigma(E)$ is the cross section of the reaction considered (i.e., d-d and d-$^3$He) and $\frac{dN}{dE}$ is the distribution function (i.e., MB, sMB or LN distributions). In Fig. 4, a typical situation is displayed, where both the Maxwell-Boltzmann, the shifted MB and the log-normal distributions reproduce the measured signals correctly. The only substantial differences among them are in the high momentum tail of the signal. As mentioned above, these deviations become more dominant when calculating the fusion yield because the average cross section in equation (15) is the convolution between the cross section, which increases exponentially at higher energies, and the distribution function which decreases for increasing energies. The log-normal distribution function indeed, decreases slower at higher energy because of the natural logarithm dependence in the exponent, which results in an overestimate of the fusion yields. Of course this can be corrected by making opportune adjustment to the log-normal distribution, building the distribution (LN2) so that we are able to reproduce the fusion yield (see Fig. 4, dash-dot orange line). In this case though, the adjusted distribution does not reproduce the experimentally measured high energy ion spectrum. The quantities $\frac{dY_{n(BB)}}{dE} = \rho\sigma(E)l\frac{dN}{dE}$ and $\frac{dY_p}{dE} = \rho_{^3He}\sigma(E)l\frac{dN}{dE}$ are plotted in Fig. 5 for the different distribution functions and the experimental data. The fusion cross sections ($\sigma(E)$) are taken from Ref. [44].

In Fig. 6, we plot the number of fusions obtained in different shots and compared to the estimates from the LN distributions (top panels) and the Maxwell-Boltzmann and the shifted MB distributions (bottom panels). The convolution of the distribution function with the fusion cross sections exhibits a maximum which is usually referred as Gamow energy $E_G$ [45, 46]. Such a quantity could be directly determined from the data only for the d-d case [36] (see Fig. 5) because of the noise.

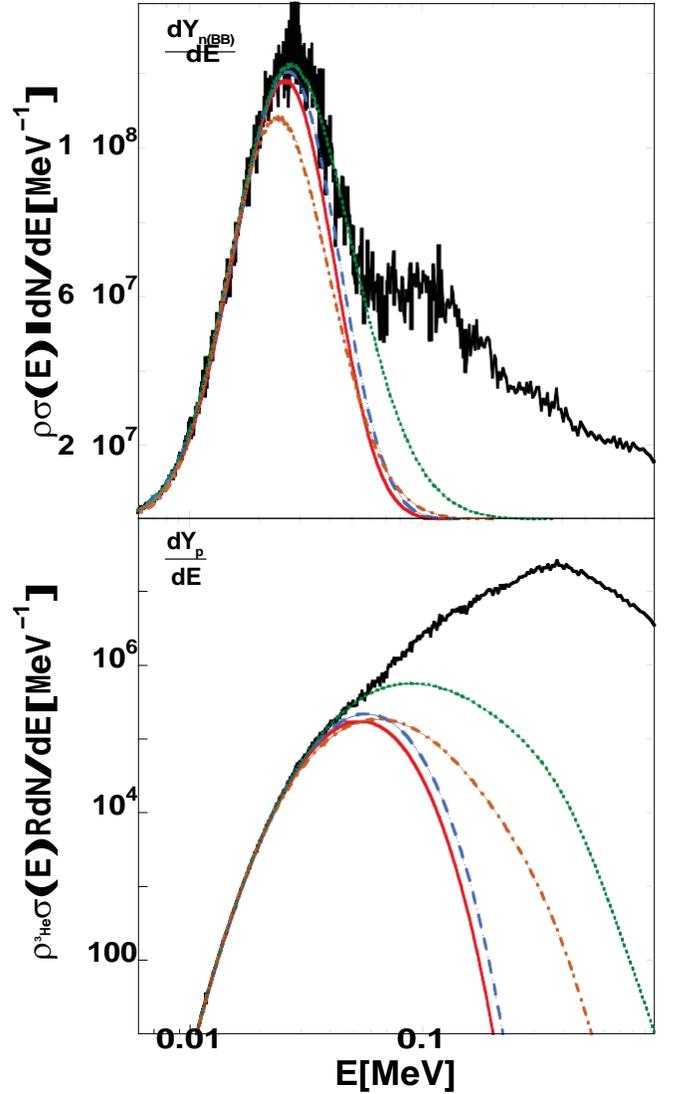

FIG. 5. The integral of the quantities displayed in the figure gives the fusion yield obtained for d-d from the hot deuterium ions contribution ($Y_{n(BB)}$, top panel) and d-$^3$He ($Y_p$, bottom panel) fusion reactions, respectively. Maxwell-Boltzmann distribution (solid red line), shifted Maxwell-Boltzmann distribution (dashed blue line) and log-normal distribution (dash-dot orange line) all give the correct measured fusion yields within the error bars. The log-normal distribution in green, which describe the ion signal correctly, does not reproduce the number of fusions measured, especially d-$^3$He which is more sensitive to the highest energies. A more precise measurement of the fusion yields, i.e. with smaller error bars, might distinguish further among the different distributions.

We can easily estimate it for each theoretical distribution and for both nuclear reactions. The calculated number of fusions as function of the Gamow energy is also given in Fig. 6 (right panels). Note that the d-$^3$He results in a higher Gamow energy because of the higher Coulomb barrier, thus it is more sensitive than d-d reactions to higher energy plasma ions, which reaffirms the importance of accurate measurement of the high en-ergy ions. Comparing the Gamow energies for the different distributions, we find that the MB distributions give similar values, while the LN distributions give generally higher values because of the slow energy decay of the distribution. In the particular case shown in Figs. 4 and 5 (i.e., shot #2754), the experimentally measured ion distribution energy spectrum and d-d fusion reaction yield are well described by the LN1 distribution (green dotted line). On the other hand, the d-$^3$He fusion yield we obtain with the LN1 distribution largely overestimates the data (see Fig. 6, top panel). In general for all the other cases in Fig. 6, the estimated number of fusions for d-d and d-$^3$He nuclear reactions using LN1 distributions are systematically higher than the measured fusion yields, above the experimental error. We can reproduce rather well both fusion yields of neutrons and protons produced with the LN2 distributions if we adjust the parameters, say to reproduce the number of d-d fusions. In such a case we do not reproduce the high momentum tail of the ion kinetic energy distribution, see Fig. 4.

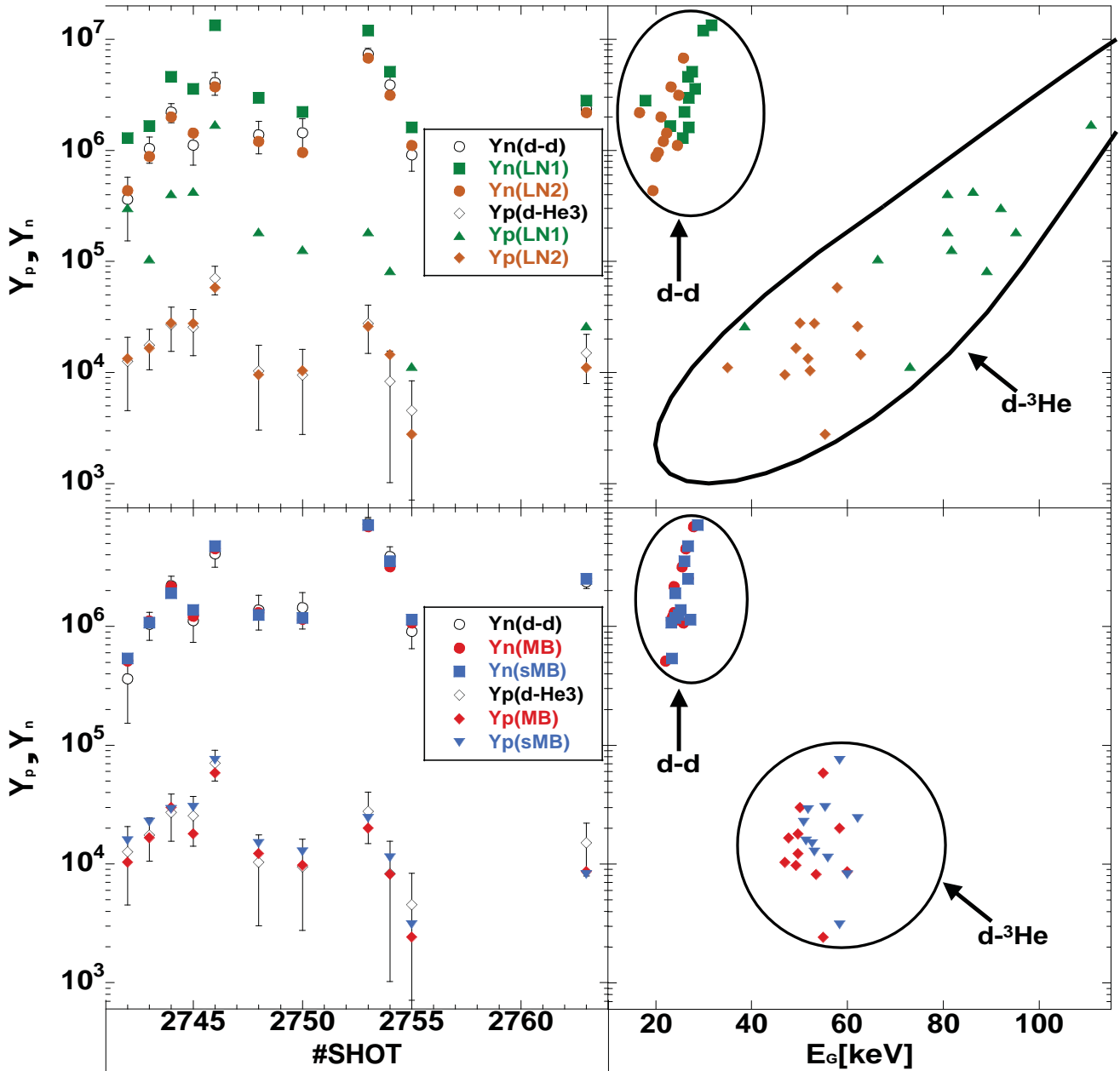

FIG. 6. Total fusion yield obtained for d-d ($Y_n$) and for d-$^3$He fusion reaction ($Y_p$). Open symbols refer to the experimentally measured fusion yields. MB and shifted MB distributions (bottom panels) all give the correct measured fusion yields within the errors. Log-normal distribution LN2 (top panels) gives the correct measured fusion yields when the parameters are chosen to reproduce d-d fusions. The log-normal distribution LN1 (top panels), which describes the ion signal correctly, does not reproduce the number of fusions measured, especially d-$^3$He which is more sensitive to the highest energies. The right panels show the estimated fusion yields as function of the Gamow energies $E_G$ obtained for each distribution. The LN1 distribution results in higher Gamow energies as expected.

## VI. CONCLUSIONS

We used the Maxwell-Boltzmann (MB) distribution, the shifted MB distribution (sMB) and the energy distribution derived from a log-normal size distribution of clusters (LN1, LN2) to estimate the fusion yields from d-d and d-$^3$He fusion reactions. We have shown that the first two distributions reproduce well the experimentally measured ion kinetic energy distributions and both fusion yields of neutrons and protons produced, offering a situation in which the ion distribution can be considered in thermal equilibrium with or without a negative collective energy. On the shots displayed in Fig. 6, we observe an average deuterium ion kinetic energy of 12.9 ± 2.3 keV, or $kT$ = 8.6 ± 1.5 keV defined as two thirds of the average kinetic energy of deuterium ions, compared with an average deuterium ion kinetic energy of 14.3 ± 2.4 keV, $kT$ = 9.5 ± 1.6 keV and $E_C \sim$ -11 keV using shifted MB distributions. These small differences could be further resolved with a better precision in the measurements of the ion kinetic energy distributions and the fusion yields. In contrast, the LN distribution does not reproduce both the measured ion signal and the fusion particle yields with the same set of parameters within the experimental errors. In fact, we were able to derive the correct fusion yield (LN2), but in this case the log-normal distribution does not give a good representation of the measured ion signal. However, as pointed out in Ref. [32], the measured deuterium cluster sizes are far smaller than the average sizes calculated using the Hagena parameter [47, 48] which might have affected the cluster size distribution itself, and therefore made our description based on [3, 25, 37, 38, 41, 42] and the LN distribution not ideal. In any case, at least for the quantities considered here, we conclude that the distribution is identical to that expected for thermal equilibrium, perhaps because the Coulomb explosion of different cluster sizes is so chaotic to be practically indistinguishable from thermalization [33, 40]. The approach discussed in this work could be very useful to describe astrophysical scenarios reproduced in laboratory, especially for plasma fusion cross sections measurement in the presence of electrons.

## ACKNOWLEDGMENTS

The experimental work was done at the University of Texas at Austin and was supported by NNSA Cooperative Agreement No. DE-FC52-08NA28512 and the DOE Office of Basic Energy Sciences. The analysis of the data was performed at the Texas A&M University and was supported by the U.S. Department of Energy, Office of Science, Office of Nuclear Physics, under Award No. DE-FG03-93ER40773 and by the Robert A. Welch Foundation under Grant No. A0330. W.B. was supported by the Los Alamos National Laboratory LDRD program. M. W. was supported by the National Science Foundation Graduate Research Fellowship under Grant No. 1263281.